\documentclass{aa}
\usepackage{psfig}
\begin{document}

\thesaurus{03(11.01.2; 11.19.1; 11.09.1 NGC 7582)}

\title{The 10-200 $\mu$m spectral energy distribution of the prototype 
Narrow-Line X-ray galaxy NGC 7582
\thanks{Based on observations with ISO, an ESA project with instruments funded 
by ESA Member States (especially the Principal Investigator countries: France,
Germany, Netherlands and the United Kingdom) and with the participation of 
ISAS and NASA.}
}

\subtitle{}

\author
{Mario Radovich\inst{1}, Ulrich Klaas\inst{1}, Jos\'e Acosta-Pulido\inst{2}  
\and Dietrich Lemke\inst{1}}
\institute
{Max-Planck-Institut f\"ur Astronomie, K\"onigstuhl 17, D-69117 Heidelberg,
Germany\\ \and Instituto de Astrofisica de Canarias, E-38200 La Laguna, 
Tenerife, Spain}
\authorrunning{M. Radovich et al.}
\titlerunning{The 10-200 $\mu$m SED of NGC 7582}
\offprints{M. Radovich (radovich@mpia-hd.mpg.de)}

\date{Received 16 March 1999 /Accepted 7 June 1999}

\maketitle

\begin{abstract}
We present the spectral energy distribution (SED) between 10 and 200 $\mu$m 
obtained for the prototype Narrow-Line X-Ray Galaxy NGC 7582 with ISOPHOT, the 
photometer on board the {\em Infrared Space Observatory}. 
The emission is spatially extended and we separated for the first time
the nuclear and extranuclear infrared SEDs.   
The nuclear luminosity (L$_{\rm IR} \sim 4.5\times10^{10}$ $L_\odot$) is 
dominated by cold ($T \sim$ 32 K) dust emission mainly due to star formation 
activity; 
warm ($T \sim$ 122 K) dust emission is also present and is probably related to 
the active nucleus. In addition to a cold component of 30 K, the extranuclear 
SED (L$_{\rm IR} \sim 1.7\times10^{10}$ $L_\odot$) shows emission by colder 
($T \sim$ 17 K) dust:  this very cold component comprises 90\% of the total 
dust mass of $9.8\times10^7$ $M_\odot$.

\end{abstract}

\section{Introduction}

\object{NGC 7582} (z=0.00525) is a highly inclined  ($i \sim 60 \degr$) barred spiral 
galaxy; its blue luminosity corrected for galactic and internal extinction 
is $\sim3.5\times10^{10}$ L$_\odot$ 
(Claussen \& Sahai \cite{claussen}) with  a Hubble constant  
H$_0$ = 75 km s$^{-1}$ Mpc$^{-1}$. 
It was classified by Mushotzky (\cite{musho})
as a Narrow-Line X-ray Galaxy (NLXG) due to its high  X-ray luminosity,  
L(2-10 keV) $\sim 7\times10^8$ L$_\odot$, typical of low-luminosity Seyfert 1 galaxies,
but with narrow (FWHM $<$ 200 km s$^{-1}$) emission lines, 
more typical of Seyfert 2 or starburst galaxies.

\begin{figure}
\psfig{file=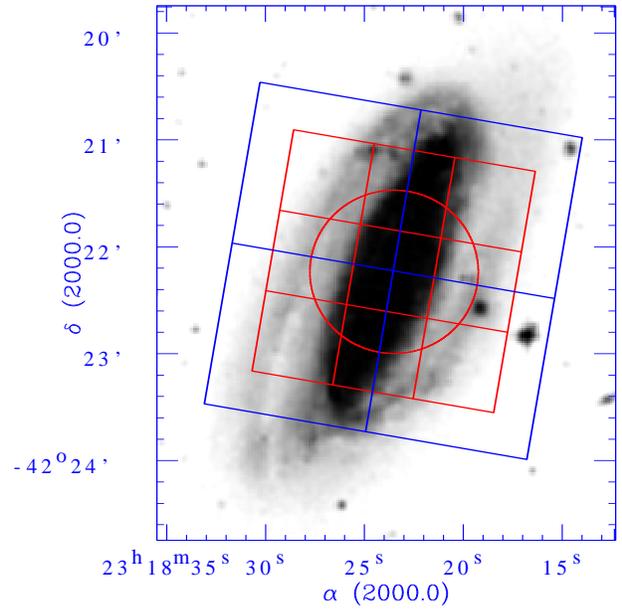,bbllx=70pt,bblly=65pt,bburx=530pt,bbury=520pt,width=8.3cm,angle=270}
\caption{The positions of the circular PHT-P aperture 
and of the  C100 (3x3) and C200 (2x2) arrays are overplotted on a J-band image of NGC 7582
obtained from the SkyView survey analysis system.}
\label{fig:image}
\end{figure}

\begin{figure}
\psfig{file=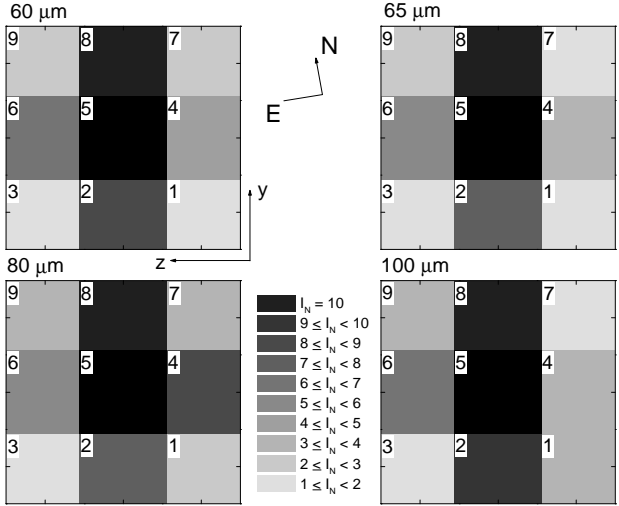,bbllx=40pt,bblly=90pt,bburx=590pt,bbury=550pt,width=8.3cm,angle=0,clip=}
\caption{Intensity distribution  in the C100 array after background subtraction. 
Intensities in the border pixels have been normalized to the range 1-10, with darker
colours indicating increasing intensities; the intensity in the central pixel is outside  
the scale.}
\label{fig:3dc100}
\end{figure}

\begin{figure}
\psfig{file=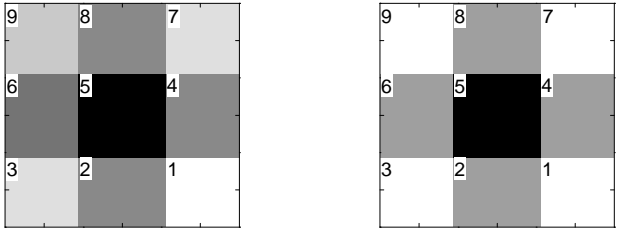,bbllx=40pt,bblly=315pt,bburx=590pt,bbury=525pt,width=8.3cm,angle=0,clip=}
\caption{Intensity distribution produced by a pointlike source in the C100 array.  The intensity 
in the central pixel and the intensity scale are the same as those in the 60$\mu$m filter
 for NGC 7582, displayed in Fig.~2.
{\em Left: } observations of the star HR 5340; {\em right: } 
theoretical monochromatic footprint as given in Table~2. }
\label{fig:hr5340}
\end{figure}

Recent ASCA observations (Schachter et al. \cite{schachter}, Xue et al. 
\cite{xue}) revealed high variability on short ($\sim$ 5.5 hours) time scales 
in the hard X-ray range (2-10 keV), typical of an AGN; in particular, 
Schachter et al. (\cite{schachter}) report a hard X-ray flux drop by 40\% in 
$\sim$ 6 ks, implying a source size of $\sim 2\times10^{14}$ cm. 
In addition, the existence of an active
nucleus is suggested by the detection  of an ionization cone in [\ion{O}{III}] 
$\lambda$5007 narrow--band images (Storchi-Bergmann \& Bonatto \cite{storchi})
and by the presence  of high ionization emission lines 
(e.g. [\ion{Ne}{V}] $\lambda$3426).
Genzel et al. (\cite{genzel}) report the detection  of both high-ionization   
(e.g. [\ion{O}{IV}] $\lambda$25.9$\mu$m, [\ion{N}{V}] $\lambda$14.3$\mu$m) 
and low--ionization (e.g. [\ion{Ne}{II}] $\lambda$12.8$\mu$m, 
[\ion{S}{III}] $\lambda$18$\mu$m) emission lines with the ISO short-wavelength
grating spectrometer (SWS). The low-ionization lines may be produced either by 
a starburst or by an AGN  (Radovich et al. \cite{radovich}).
A composite model, where high-ionization lines  are produced in gas 
ionized by the AGN and  the low-ionization lines are enhanced in starburst 
regions, is also possible and would explain the 
[\ion{O}{III}] $\lambda$5007/H$\beta$ ratio of 
$\sim$ 3 (Storchi-Bergmann et al. \cite{storchi-kinney}), lower than usually 
observed in AGNs.  
According to the `unified model' (Antonucci \cite{antonucci}), 
broad emission lines are extinguished in Seyfert 2 galaxies by a
dusty torus located at $\le$ 100 pc from the nucleus, but they should be 
scattered into our line of sight  by free electrons around the torus and 
should be observed in polarized light. However, Heisler et al. 
(\cite{heisler}) did not detect broad  emission lines in the polarized 
spectrum of NGC 7582: they attributed the non-detection of  polarized broad 
lines to the fact that our line of sight passes through an amount of dust  
in the torus which is larger than in Seyfert 2s with polarized broad lines.

The presence of strong star formation activity in the central kpc was 
concluded from other indicators. Morris et al. (\cite{morris}) showed 
that in the inner kpc the H$\alpha$ emission suggests a rotating disk 
of HII regions in the plane of the galaxy. Recently Aretxaga et al. 
(\cite{aretxaga}) reported the appearance of broad (FWHM $\sim$ 10000 km 
s$^{-1}$) 
permitted lines in the period July - October 1998: they explained it as 
the result of supernova explosions in
the circumnuclear starburst, rather than as a change of the reddening in the 
torus which would allow to see the inner nuclear regions.
The non-thermal radio emission probably
originates in many supernova remnants associated with this
starburst, as found by Forbes \& Norris (\cite{forbes}).
In the infrared, the 8-13 $\mu$m emission is characterized by the
presence of polycyclic aromatic hydrocarbon (PAH) features as revealed by 
ground-based  (Frogel et al. \cite{frogel}, Roche et al. \cite{roche}) and,
more recently, by ISO SWS observations (Genzel et al. \cite{genzel}). Their
strength is much more typical of starbursts rather than of AGNs: in fact, 
Genzel et al. (\cite{genzel}) found that the typical strength of the 
7.7$\mu$m PAH feature, defined in their Table 1, is 0.04 in AGNs and 3.6 
in starbursts; they find a value of  2.5 in NGC 7582.
Multi-aperture observations performed by Frogel et al. 
(\cite{frogel}) showed that the 10 $\mu$m emission comes from a region smaller
than 8$\arcsec$, but its origin is still unclear. 
Using the FIR to X-ray luminosity correlation for normal and starburst galaxies, 
Turner et al. (\cite{turner}) found that the maximum contribution of a 
starburst to the 0.5-4.5 keV emission is $\sim 34\%$.
According to Xue et al. (\cite{xue}), the soft X-ray emission (0.5 - 2 keV) 
is the sum of scattered emission from the nucleus and emission
from the starburst, the latter being $\sim$ 10-20\%. 
In their Hubble Space Telescope imaging survey of nearby AGNs, 
Malkan et al. (\cite{malkan}) detected dust lanes running across the nucleus
of NGC 7582, at a distance of some hundreds of parsecs: this dust component external 
to the torus may contribute to obliterate the broad polarized lines whereas the 
scattered X-ray photons may still be observed.

There is therefore strong evidence that NGC 7582 contains  a `buried'
AGN with the active nucleus residing in a dusty environment and 
coexisting with a circumnuclear starburst.  
Schachter et al. (\cite{schachter}) suggested that all NLXGs are
obscured AGNs.
Because of its prototype nature and brightness we observed the infrared SED of
this object with ISO (Kessler et al. 
\cite{kessler})  to determine the relative contribution of the various 
components to the total energy budget.

\begin{table}
\caption{Pointing of the observations. The on-position was centered on the
object, the off-position was used for background subtraction.}
\label{tab:log}
\begin{flushleft} 
\begin{tabular}{lccc}
\hline 
 Name & Date & RA (2000) & DEC (2000) \\ 
\hline 
 NGC 7582 on    &  24 Dec 97 &  23$^{\rm h}$18$^{\rm m}$23\fs5 & 
   -42\degr22\arcmin14\arcsec \\ 
 NGC 7582 off & 24 Dec 97 &  23$^{\rm h}$17$^{\rm m}$55\fs1 & 
  -42\degr20\arcmin41\arcsec \\
\hline
\end{tabular} 
\end {flushleft}
\end{table}

\begin{table}
\caption{Footprint values adopted for the spatial decomposition of C100 
(Herbstmeier, private communication) and C200 (Klaas et al., in preparation) data.
For C100 these are calculated values from a convolution of the monochromatic 
point spread function with the C100 pixel field size. For C200 they are empirically 
derived from a comparison of measurements of point sources centered on the arrays
with raster measurements centered on each pixel.}
\label{tab:fp}
\begin{flushleft} 
\begin{tabular}{ccccc}
\hline 
\multicolumn{3}{l}{C100 Detector} \\
 Pixel & 60$\mu$m & 65$\mu$m &  80$\mu$m & 100$\mu$m \\
1 & 0.0087 & 0.0132 & 0.0178 & 0.0228 \\
2 & 0.0429 & 0.0423 & 0.0379 & 0.0385 \\
3 & 0.0087 & 0.0123 & 0.0176 & 0.0228 \\
4 & 0.0440 & 0.0434 & 0.0377 & 0.0382 \\
5 & 0.6625 & 0.6587 & 0.6416 & 0.5561 \\
6 & 0.0440 & 0.0434 & 0.0377 & 0.0382 \\
7 & 0.0087 & 0.0123 & 0.0176 & 0.0228 \\
8 & 0.0429 & 0.0423 & 0.0379 & 0.0385 \\
9 & 0.0087 & 0.0132 & 0.0178 & 0.0228 \\
\hline
\multicolumn{3}{l}{C200 Detector}\\
Pixel  & 120$\mu$m & 150$\mu$m & 180$\mu$m & 200$\mu$m  \\
1-4 & 0.1435 & 0.1670 & 0.1875 & 0.1875 \\
\hline
\end{tabular} 
\end {flushleft}
\end{table}

\section{Observations and Data Reduction}
\label{sec-obsred}
NGC 7582 was observed with ISOPHOT (Lemke et al. \cite{lemke}) in staring mode 
with the multi-filter AOTs (Astronomical Observation Templates) PHT03 (PHT-P) and 
PHT22 (PHT-C) (Klaas et al. \cite{ISOman}) on the target and on one off-position 
(Table~\ref{tab:log}). 
The aperture size used with PHT-P was 99\arcsec. On-source integration times were 
64s with PHT-P and 32s with PHT-C; the same times were used for the off-source measurement.
The data reduction was performed using the ISOPHOT Interactive Analysis tool 
(PIA, version 7.2) together with the calibration data set V~4.0 
(Laureijs et al. \cite{laureijs}): corrections were made for non-linearity
effects  of the electronics, disturbancies by cosmic rays 
(deglitching) and signal dependence on the reset  interval time. The
flux calibration is based on measurements with the
thermal fine calibration sources (FCS) on board. There was one FCS measurement 
per detector in the 12 $\mu$m, 25 $\mu$m, 100 $\mu$m and 150 $\mu$m filters .
The photometric accuracies for this observing mode and brightness
range are $\sim$ 10\% in the absolute  calibration (FCS in the same filter)
and  $\le$ 30\%  in the relative  calibration (no FCS in this filter), see
Klaas et al. (\cite{IsoAcc}).

\begin{figure*}
\psfig{file=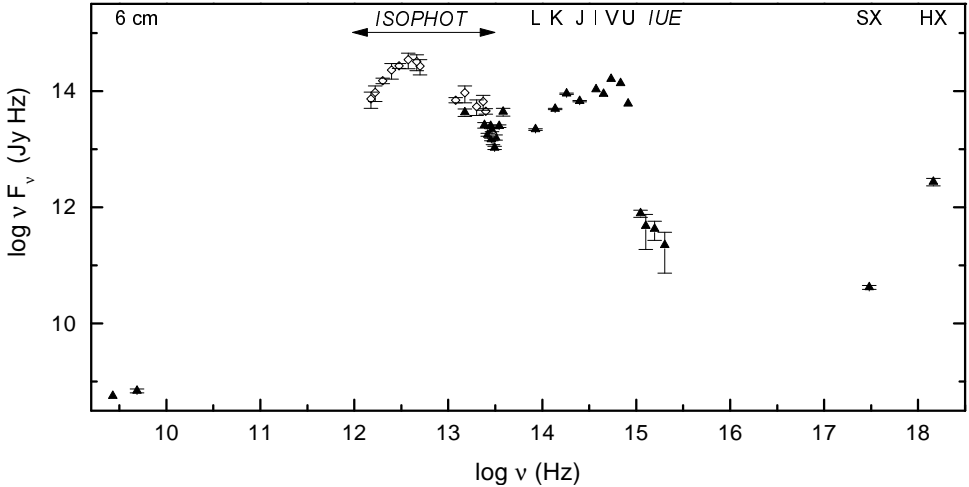,bbllx=40pt,bblly=45pt,bburx=760pt,bbury=415pt,width=13cm,angle=0,clip=}
\caption{The radio to X-ray spectral energy distribution of NGC 7582. 
Diamonds: ISOPHOT data (nucleus+disk); triangles: photometric data from the 
NASA-IPAC extragalactic database (NED), the SX (0.5-2 keV) and HX (2-10 keV)
fluxes are from Xue et al. (\cite{xue}). Uncertainties are plotted if available.}
 \label{fig:totsed}
\end{figure*}

\section{Results and Discussion}
\label{sec-results}

\subsection{Spatial decomposition}

The redshift of NGC 7582 gives a projected linear
 scale of 102 pc/arcsec. 
The diameter at the 25 mag arcsec$^{-2}$ blue isophote is 5\arcmin\
(De Vaucouleurs et al. \cite{devauc}), which is larger than both the
C100 and C200 detectors ($\sim$ 2\farcm2 and 3\arcmin, respectively, see
 Fig.~\ref{fig:image}). Indeed, a first hint that infrared emission 
  is extended was found by Sanders et al. (\cite{sanders})
  in the IRAS Bright Galaxy Sample data reprocessed using the HIRES algorithm. 
  They report extended emission in NGC 7582 at 12 and 25 $\mu$m (size $> 0\farcm75$) 
 and 60 $\mu$m (size $> 1\farcm75$), but could not resolve it at 100 $\mu$m 
(size $< 2\farcm8$). 
The intensity distribution in the C100 array after background subtraction is 
displayed in Fig.~\ref{fig:3dc100}. For comparison we show in Fig.~\ref{fig:hr5340} the
patterns expected from a pointlike source. The {\em left} pattern is that 
{\em observed} for a star, the {\em right} pattern was {\em generated} using 
the calculated monochromatic footprint values of Table~\ref{tab:fp};
in both cases the intensity in the central pixel was matched to that found in 
NGC 7582 (60$\mu$m filter). 
The two distributions are slightly different since the observed one includes effects by small telescope
pointing offsets and chromatic effects within the width of the bandpass of the 60$\mu$m filter.
It can be seen that the intensity distribution in NGC 7582 is not that expected from a pointlike
source. Pixels 2 and 8, which are aligned along the major axis of the galaxy, clearly show an 
excess emission;  the values in the other border pixels are closer to those given by
 the intensity distribution of a pointlike source. 

In order to disentangle the extended and pointlike components, we proceeded as follows.  
We assumed that ({\em i}) the emission is composed of a compact 
central source plus extended emission and ({\em ii}) the array is centered on 
the compact source.  
The total intensity observed in each pixel is then given by: 
 
\begin{equation}
\label{eq:psf}
I_{\rm i} = f_{\rm i} I_{\rm nuc} + I_{\rm e, i},
\end{equation}

where $f_{\rm i}$ is the fraction of the intensity from the pointlike source 
falling on each pixel (Table~\ref{tab:fp});
I$_{\rm nuc}$ is the intensity emitted by the pointlike {\em nuclear}
source, I$_{\rm e,i}$ is the extended emission from the extranuclear
regions detected by each pixel. The total extranuclear emission is defined as
$I_{\rm e} = \sum_{i=1}^{\rm np}{I_{\rm e,i}}$ (C100: np=9, C200: np=4): we shall 
call it the {\em disk} emission hereafter. For C100 we made the further assumption
that $I_{\rm e, 5}$ is the same as in pixels 2 and 8, $I_{\rm e, 5} = \left(I_{\rm e, 2} +
I_{\rm e, 8}\right)/2$; this seems plausible from the observed intensity distribution on 
the array (Fig.~\ref{fig:3dc100}). 

For C200 this approach was not feasible due to the lack of a central pixel:  
we extrapolated the nuclear flux ($I_{\rm nuc}^{\ast}$) from the blackbody 
fit of the decomposed nuclear C100 values. The extranuclear component was then obtained
using Eq.~\ref{eq:psf} with the footprint scaling for C200 (Table~\ref{tab:fp}): 

\begin{equation}
\label{eq:psfc2}
I_{\rm e, i} = I_{\rm i} - f_{\rm i} I_{\rm nuc}^{\ast}.
\end{equation}

An estimate of the uncertainties introduced by this procedure was computed by propagating the 
statistical uncertainty of the intensities in each pixel, which is typically
$\Delta I_{\rm i}/I_{\rm i} < 5\%$ (Laurejis \& Klaas \cite{spd_errb}). The comparison 
(see Table~\ref{tab:fluxes}) with the 
photometric accuracies quoted in  Sect.~\ref{sec-obsred} shows however, that the final uncertainties are
dominated by systematic errors in background subtraction and relative filter-to-filter calibration 
which affect the total measurement, rather than the individual pixel measurements.

\subsection{Spectral Energy Distribution} 
In Fig.~\ref{fig:totsed} we show the SED of NGC 7582 from the radio to  
X--ray frequencies, including the new ISOPHOT data points. 
The radio to near-IR SEDs are similar in Seyfert and starburst galaxies 
(Schmitt et al. \cite{schmitta}). 
The peak of the SED is located in 
the far-infrared and is likely due to the dust heated by the starburst. The rise 
in the blue wavelengths of the visual range can be either due to the AGN or 
to the starburst. Schmitt et al. (\cite{schmittb}) found that nearly 50\% of 
the flux at $\lambda$5870 \AA\ is produced by a population of stars with an 
age $\le$ 100 Myr, and that the contribution from the nuclear featureless continuum 
is only $\le 5\%$; the remaining 45\% comes from stars older than 1 Gyr. 
The ultraviolet emission was detected by IUE in a $10\arcsec\times20\arcsec$
aperture (linear projected size of 1 kpc $\times $2 kpc). Heckman et al. 
(\cite{heckman}) suggested that the steep ultraviolet continuum, 
$f_\lambda \propto \lambda^{1.7}$, may be produced by a very dusty starburst. 
This was later confirmed by Bonatto et al. (\cite{bonatto}), who reproduced 
the ultraviolet continuum of a sample of nearby spiral galaxies 
by stellar population synthesis and concluded that in the case of NGC 7582
the UV light  is dominated by recent star formation with an age $t \le 500$ Myr
in a reddened nuclear starburst.
The presence of the active nucleus becomes evident only in the hard X--ray 
domain.

\begin{figure*}
\psfig{file=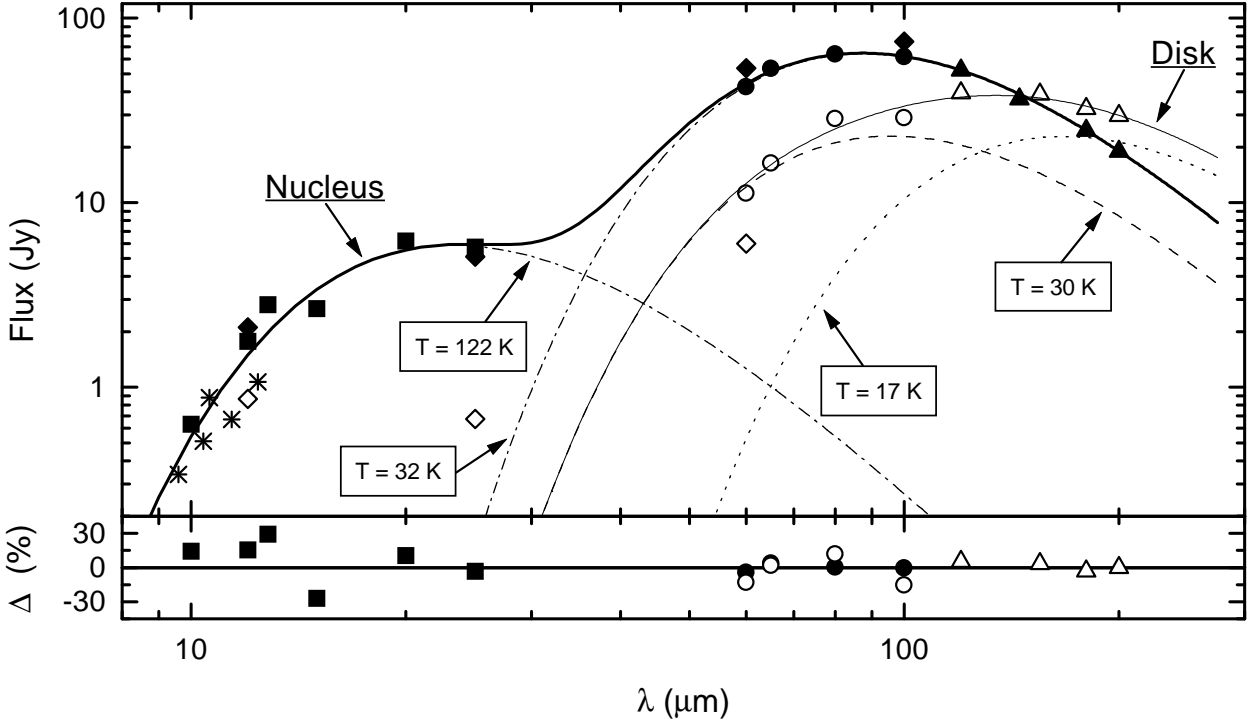,bbllx=50pt,bblly=35pt,bburx=760pt,bbury=450pt,height=10cm,angle=0}
\caption{Decomposition of the SED of NGC 7582.
Filled symbols: nuclear fluxes; open symbols: disk fluxes. 
Squares: PHT-P; circles: C100; triangles:
C200 (the 150 $\mu$m disk and nuclear points have been slightly shifted in 
wavelength to avoid overlapping). Diamonds: IRAS data from Sanders et al. 
(\cite{sanders}). Asterisks: ground-based observations from Roche et al. 
(1984). 
The black body fits are indicated as follows. Nucleus: dashed-dotted
lines (PHT-P and PHT-C components); the thick line gives the sum of the two 
components. Disk: dashed (C100) and dotted (C200) line; the thin line shows 
the sum of the two components.
The lower panel displays the deviation of the measured ISOPHOT photometric 
data from the fit.}
 \label{fig:sed}
\end{figure*}

The ISOPHOT spectral energy distribution and its decomposition into several spatial 
and temperature components is presented in Fig.~\ref{fig:sed}. 
We fitted  them with a modified blackbody, i.e. a blackbody with a wavelength dependent
emissivity (see e.g. Hildebrand \cite{hilde}); we adopted an emissivity $\propto \lambda^{-2}$. 
Colour correction was performed in an iterative way, namely a first value of the 
temperature was adopted and the corresponding colour correction factors applied: the
resulting SED was fitted with modified blackbodies using a non-linear least
squares algorithm giving new values of the temperature. The whole procedure was
then repeated until convergence. The fluxes derived  are listed in 
Table~\ref{tab:fluxes}, the temperature components are given in 
Table~\ref{tab:lums}.
The fluxes measured in the PHT-P filters ($\le$ 25 $\mu$m) with the 99\arcsec\ 
aperture are in good agreement with the values of 
ground-based observations (Roche et al. \cite{roche}) with smaller 
apertures (8\arcsec):
we therefore conclude that an extranuclear component is negligible.
The 12 and 25 $\mu$m ISO fluxes are also in good agreement with the peak IRAS 
intensities  given by Sanders et al. (\cite{sanders}) with an angular 
resolution (width at 25\% peak intensity $\sim$ 94\arcsec\ and 70\arcsec, respectively) 
comparable  to that of the PHT-P aperture.

\begin{table*}
\caption{Central wavelengths and spectral resolutions of the ISOPHOT filter bands, 
total fluxes (not colour corrected) 
and fluxes obtained after separation of the nuclear and disk components, blackbody 
fitting and colour correction; statistical uncertainties for the decomposed components (in italics)
 and  photometric accuracies are also given. 
We show for comparison the IRAS fluxes from Sanders et al. (\cite{sanders}); a proper 
colour correction is applied only to the peak and extended values. Fluxes and 
uncertainties are in Jy.}
\label{tab:fluxes}
\begin{flushleft}
\begin{tabular}{@{\extracolsep{.1cm}}cc cccccccc c cccc}
 \hline
\noalign{\smallskip}
              &                                   &    \multicolumn {8}{c}{ISOPHOT}                                            &  & \multicolumn {3}{c}{IRAS}    \\
\cline{3-10} \cline{12-15}
$\lambda_c$   &    $\frac{\lambda_c}{\Delta\lambda_c}$&    Total & phot  &  Nucleus  & {\em stat} & 
phot &  Disk & {\em stat} & phot & & Total  &    &  Peak & Extended\\

\multicolumn{4}{l}{P1 Detector}       \\                                                    
10  &  5.6  &  0.6  & $\pm$0.2  &  0.6  &    &  $\pm$0.2  &    &    &    &  &          &    &    \\
12  &  5.6  &  1.5  &  $\pm$0.2  &  1.8  &    &  $\pm$0.2  &    &    &    &  &  2.3  &  $\pm$0.1    &  2.1  &  0.9  \\
12.8  &  5.3  &  2.8  &  $\pm$0.8  &  2.8  &    &  $\pm$0.8  &    &    &    &  &          &    &    \\
15  &  1.8  &  2.6  &  $\pm$0.8  &  2.7  &    &  $\pm$0.8  &    &    &    &  &          &    &    \\
\multicolumn{4}{l}{P2 Detector}       \\                                                    
20  &  2.3  &  6  &  $\pm$2  &  6  &    &   $\pm$2 &    &    &    &  &          &    &    \\
25  &  2.6  &  7.2  &  $\pm$0.7  &  5.8  &    &  $\pm$0.6  &    &    &    &  &  7.5  &  $\pm$0.4    &  5.1  &  0.7  \\
\multicolumn{4}{l}{C100 Detector}       \\                                                     
60  &  2.5  &  57  &  $\pm$17  &  42  & {\em $\pm$ 2  } &  $\pm$13  &  11  & {\em  $\pm$1  } &  $\pm$3  &  &  52  &  $\pm$3    &  54  &  6  \\
65  &  1.1  &  59  &  $\pm$18  &  53  & {\em $\pm$ 3  } &  $\pm$16  &  16  & {\em  $\pm$1  } &  $\pm$5  &  &          &    &    \\
80  &  1.7  &  81  &  $\pm$24  &  64  & {\em  $\pm$4  } &  $\pm$19  &  28  & {\em  $\pm$2  } &  $\pm$9  &  &          &    &    \\
100  &  2.2  &  85  & $\pm$9  &  62  & {\em  $\pm$4  } &   $\pm$6  &  29  & {\em  $\pm$2  } &  $\pm$3  &  &  78  &  $\pm$5    &     &     \\
\multicolumn{4}{l}{C200 Detector}       \\                                                     
120  &  2.4  &  90  &  $\pm$27  &  52  & {\em  $\pm$3  } &  $\pm$16  &  39  & {\em  $\pm$4  } &  $\pm$12  &  &           &     &     \\
150  &  1.9  &  75  &  $\pm$8  &  36  & {\em  $\pm$2  } &  $\pm$4  &  39  & {\em  $\pm$4  } &  $\pm$4  &  &          &    &    \\
180  &  2.6  &  59  & $\pm$18  &  25  & {\em  $\pm$1  } &  $\pm$7  &  32  & {\em  $\pm$3  } & $\pm$10  &  &          &    &    \\
200  &  3.6  &  50  &  $\pm$15  &  19  & {\em  $\pm$1  } &  $\pm$6  &  30  & {\em  $\pm$2  } &  $\pm$9  &  &          &    &    \\
\hline
\end{tabular}
\end{flushleft}
\end{table*}

\begin{table*}
\caption{Aperture diameters and derived projected linear sizes (radii), infrared 
luminosities, dust masses and star formation rates derived for the different
components; uncertainties given by the blackbody fit are displayed.}
\label{tab:lums}
\begin{flushleft}
\begin{tabular}{c c c c c c}
\hline
Aperture & Radius & T  & L(1-1000 $\mu$m) & M$_{\rm dust}$ & SFR \\
$[\arcsec]$ & $[$kpc] & [K] & $[L_\odot]$  & $[M_\odot]$ & 
$[M_\odot$ yr$^{-1}]$ \\
\hline
\multicolumn{2}{l}{Nucleus}\\
99 & $<$ 5  & 122$\pm$5 & $(12^{+5}_{-4})\times10^{9}$ & $(11^{+2}_{-1})\times10^{2}$ & -- \\
43 & $<$ 2 & 32$\pm$0.3  & $(33^{+4}_{-4})\times10^{9}$  & $(93^{+5}_{-5})\times10^{5}$ &
 $5.8^{+0.7}_{-0.6}$ \\
\multicolumn{2}{l}{Disk}\\
138 & $\le$ 7 & 30$\pm$2 & $(11^{+8}_{-5})\times10^{9}$  & $(5^{+1}_{-1})\times10^{6}$ & 
$2^{+1}_{-1}$ \\
182 & $\le$ 9 & 17$\pm$3 & $(6^{+10}_{-5})\times10^{9}$ & $(8^{+2}_{-2})\times10^{7}$ & -- \\
\hline
\end{tabular}
\end{flushleft}
\end{table*}

The results of the decomposition of the nuclear and disk SEDs are:

\begin{description}
\item{{\em 1. Nucleus} --} The SED between 10 and 25$\mu$m (PHT-P) is well 
fitted by a warm ($T = 122$ K) component which is typical of Seyfert galaxies 
(P\'erez Garc\'{\i}a et al. \cite{perez}), where dust in the torus is 
heated by the active nucleus. 
 The deviations of the measured data from the fit are within 
the general calibration accuracy, $\Delta \le$ 15\%, with the exception of the 12.8 
$\mu$m and 15 $\mu$m values ($\Delta \sim$ 30\%). The contribution of the [\ion{Ne}{II}] 
$\lambda$12.8 $\mu$m line cannot account for the excess in the 12.8 $\mu$m band: 
in fact the intensity given by Genzel et al. (\cite{genzel}) amounts only to 
$\sim$ 3\% of the integrated flux in this band.
At $\lambda > 60\mu$m (C100) we see a well defined cold (T $\sim$ 32 K) 
component. Since we 
have $S_\nu(25)/S_\nu(60) = 0.13$, whereas in Seyfert galaxies $S_\nu(25)/S_\nu(60) > 0.2$
(Miley et al. \cite{miley}), we conclude that at these wavelengths the emission is 
dominated by dust heated by the starburst. According to the colour-colour diagrams 
by Dopita et al. (\cite{dopita}), the FIR colours of NGC 7582 may be interpreted
as the composition of substantial circumnuclear star formation emission plus 
emission from an active nucleus observed through a dusty environment  
($\tau_{12 \mu m} > 3$): the dust may be located both in the accretion torus and 
in the circumnuclear starburst.

\item{{\em 2. Disk} --}  The signal in C100 is dominated by a component whose temperature 
(T $\sim$ 30 K) is similar to that found in the nucleus.
 The signal in C200 reveals a colder component, T $\sim$ 17 K; due to absence of data 
points longward of 200 $\mu$m, the fit is not very well constrained and has an 
uncertainty $\Delta T \sim$ 3 K.
This temperature is typical of dust heated by the interstellar radiation field;
the presence of very cold dust which dominates the emission
in the outer regions of the disk has also been  found in the analysis of ISOPHOT data 
both for active  (P\'erez Garc\'{\i}a et al. \cite{perez}) and normal 
(Kr\"ugel et al. \cite{krugel},  Alton et al. \cite{alton}) galaxies. 
\end{description}

\subsection{Luminosities and dust masses}

Luminosities between 1 and 1000 $\mu$m derived  for the
different components are compiled in Table~\ref{tab:lums}.
The ratio of the nuclear starburst to disk luminosity  is $2^{+2}_{-1}$. 
Rowan-Robinson \& Crawford (\cite{rowan}) obtained a 
value of 2.1 fitting the IRAS values with starburst and disk emission models: 
this provides a  reliability check for the decomposition based on the 
spatial resolution we obtained.

The spatial and spectral decomposition allows an estimate which is more 
reliable than with IRAS of the star 
formation rates and dust masses (Table~\ref{tab:lums}). 
Star formation rates  were derived for the cold (30 K) nuclear and disk components only,
according  to Kennicutt (\cite{kenn}); dust masses were 
computed according to Klaas \& Els\"asser (\cite{klaas}), assuming dust grain properties as 
in Hildebrand (\cite{hilde}) for the emissivity  adopted here ($\propto \lambda^{-2}$): 
\begin {eqnarray}
SFR & = & 1.7 \times 10^{-10} L_{\rm IR}/L_\odot\ \ [M_\odot\ \rm{yr}^{-1}],\\
M_{\rm d} & =  &  7.9 \times  10^{-5} (T_\mathrm{K}/40)^{-6} \ L_{\rm IR}/L_\odot\ \ \ [M_\odot].
\end{eqnarray}

The nuclear star formation rate of $\sim 6 M_\odot\ \rm{yr}^{-1}$ 
is typical of bright IR galaxies (Kennicutt \cite{kenn}), and confirms 
the presence of a moderate star formation activity in the inner kpc. A lower
star formation rate, $\sim 2 M_\odot\ \rm{yr}^{-1}$, is derived for the disk. 

According to Claussen \& Sahai (\cite{claussen}) the atomic and molecular hydrogen 
masses in NGC 7582 are  M(HI) = $4.3\times10^{9}$ M$_\odot$ and 
M(H$_2$) = $4.4\times10^{9}$ M$_\odot$, respectively. The dust to gas mass ratio would be 
$\sim$ 1/600, if we considered only the warm and cold components (nucleus + disk) 
detected by C100; this would give a dust mass $M_{\rm d} \sim  1.4\times10^{7}$ M$_\odot$. 
The addition of the 17 K component revealed by C200 increases the dust mass by a factor of 7;  
the dust mass is now $M_{\rm d} \sim 9.8\times10^{7}$ M$_\odot$, the dust to gas mass
ratio is 1/90, close to the canonical value.  It should be noted that due to the strong
dependence of the dust mass on the temperature, the $\Delta T = 3$ K uncertainty may change
the dust mass of the very cold component by a factor of 3. A more detailed error calculation  
shows however that the final uncertainty on the dust mass is much lower 
(see Table~\ref{tab:lums}).
The dust mass of the nuclear warm component, probably associated with the torus,  
is $M_{\rm d} \sim 3\times10^3$  M$_\odot$.

\section{Summary}
We obtained ISOPHOT data for the prototype NLXG NGC 7582 from 10 to 200 $\mu$m
and found that its emission is extended in the 60 to 200 $\mu$m range along the major axis 
of the galaxy. 
We performed a spatial decomposition into a nuclear and a disk
component. The SED in the nucleus (r $<$ 2 kpc) amounts to 70\% of the total IR luminosity 
of $6\times10^{10}$ $L_\odot$; it was decomposed into a warm (122 K) component, 
probably related to dust heated by the active nucleus, and a cold (32 K) component 
coming from a circumnuclear starburst. The IR colours are consistent with the values  
expected for an obscured active nucleus plus circumnuclear star formation. 
We conclude that the most likely interpretation of the 
NLXG nature of NGC 7582 is that the high X--ray luminosity  is 
produced by the active nucleus, which is obscured at longer wavelengths by dust 
located in the circumnuclear regions; the heating of the dust by the starburst 
produces the far-infrared emission. The availability of the data longward of 100 $\mu$m allowed 
the detection
of a very cold (17 K) component in the SED of the disk. This very cold
component is due to dust heated by the interstellar radiation field and it dominates 
the dust mass of the galaxy; its inclusion increases the dust  mass  detected now by a factor 
of 7.  

\acknowledgements{
 
The ISOPHOT consortium is headed
by the Max-Planck-Institut f\"ur Astronomie, Heidelberg, Germany. 
The ISOPHOT development and the postoperational
phase of the ISOPHOT Data Centre of MPIA Heidelberg are funded by Deutsches 
Zentrum f\"ur Luft- und Raumfahrt  (DLR), Bonn.
PIA has been jointly developed by the ESA Astrophysics Division 
and the ISOPHOT consortium.  
This research has made use of
the NASA-IPAC extragalactic database (NED) which is operated by the Jet 
Propulsion Laboratory, Caltech, under contract with the NASA. We thank
the anonymous referees for their comments which improved this paper.
}


\end{document}